\documentclass{article}

\usepackage{amsfonts} 
\usepackage{cite}
\usepackage{epsfig}
\usepackage{newalg} 
\usepackage{times}
\usepackage{cmu-titlepage}
\usepackage{url}
\usepackage{verbatim}

\newcommand{\half}{\frac{1}{2}}

\newcommand{\qmin}{q_{\mathit{min}}}

\newcommand{\algofn}[1]{{\small \textsc{#1}}}
\newcommand{\algofontcheat}{}

\newcommand{\rankvector}[1]{\langle #1 \rangle}
\newcommand{\alloccase}[1]{\langle #1 \rangle}

\newcommand{\LB}{\textit{LB}}
\newcommand{\UB}{\textit{UB}}

\newcommand{\mysection}[1]{\section{#1}}
\newcommand{\mysubsection}[1]{\subsection{#1}}
\newcommand{\mysubsubsection}[1]{\subsubsection{#1}}

\newcommand{\precaptioncheat}{}
\newcommand{\postcaptioncheat}{}
\newcommand{\mycaption}[1]{\precaptioncheat\caption{#1}\postcaptioncheat}

\newcommand{\makeplot}[2]{
    \begin{figure}[!hbt]
        \epsfig{file=increasing_items.#1.eps,width=2.2in}
        \hfill
        \epsfig{file=increasing_agents.#1.eps,width=2.2in}
        \mycaption{\it #2}
        \label{fi:#1}
    \end{figure}
}

\newtheorem{theorem}{Theorem}
\newtheorem{prop}{Proposition}
\newtheorem{defin}{Definition}

\newcommand{\propcheat}{}

\def\QuadSpace{\vspace{0.25\baselineskip}}
\def\HalfSpace{\vspace{0.5\baselineskip}}

\def\EndProof{ \quad \vrule width 1ex height 1ex depth 0pt \newline }

\newenvironment{proof}{\QuadSpace\par\noindent{\bf Proof}:}{\EndProof\HalfSpace}

\title{
Effectiveness of preference elicitation in combinatorial auctions 
}

\trnumber{CMU-CS-02-124}
\date{March 2002}
\support{This material is based upon work supported in part by the National
Science Foundation under CAREER Award IRI-9703122, Grant IIS-9800994, ITR
IIS-0081246, and ITR IIS-0121678.}
\disclaimer{The views and conclusions contained in this document are those
of the authors and should not be interpreted as representing the official
policies, either expressed or implied, of NSF or the U.S.\ government.}

\author{
Beno\^{\i}t Hudson \and Tuomas Sandholm
}

\abstract{
Combinatorial auctions where agents can bid on bundles of items are
desirable because they allow the agents to express complementarity and
substitutability between the items.  However, expressing one's
preferences can require bidding on all bundles.  Selective incremental
preference elicitation by the auctioneer was recently proposed to
address this problem~\cite{Conen01:Preference}, but the idea was not
evaluated.  In this paper we show, experimentally and theoretically,
that automated elicitation provides a drastic benefit.  In all of the
elicitation schemes under study, as the number of items for sale
increases, the amount of information elicited is a vanishing fraction
of the information collected in traditional ``direct revelation
mechanisms'' where bidders reveal all their valuation information.
Most of the elicitation schemes also maintain the benefit as the
number of agents increases.  We develop more effective elicitation
policies for existing query types.  We also present a new query type
that takes the incremental nature of elicitation to a new level by
allowing agents to give approximate answers that are refined only on
an as-needed basis.  In the process, we present methods for evaluating
different types of elicitation policies.
} 

\begin{document}
\maketitle

\mysection{Introduction}

Combinatorial auctions, where agents can submit bids on {\em bundles}
of items, are economically efficient mechanisms for selling $k$ items
to $n$ bidders, and are attractive when the bidders' valuations on
bundles exhibit {\em complementarity} (a bundle of items is worth more
than the sum of its parts) and/or {\em substitutability} (a bundle is
worth less than the sum of its parts).  Determining the winners in
such auctions is a complex optimization problem that has recently
received considerable attention
(e.g.,~\cite{Rothkopf98:Computationally,Sandholm02:Algorithm,Fujishima99:Taming,Nisan00:Bidding,Andersson00:Integer,Sandholm01:CABOB,Tennenholtz00:Some,DeVries00:Combinatorial,Hoos01:Bidding}).
%
%

An equally important problem, which has received much less attention,
is that of bidding.  There are $2^k -1$ bundles, and each agent may
need to bid on all of them to fully express its preferences.  This can
be undesirable for any of several reasons:
determining one's valuation for any given bundle can be computationally
intractable~\cite{Sand93,Parkes99:Optimal,Larson01:Deliberation,Sandholm00:Issues,Larson01:Computationally};
%
%
there is a huge number of bundles to evaluate;
communicating the bids can incur prohibitive overhead (e.g., network
traffic); and
agents may prefer not to reveal all of their valuation information due to
reasons of privacy or long-term competitiveness~\cite{Rothkopf90:Why}.
Appropriate bidding
languages~\cite{Sandholm02:Algorithm,Fujishima99:Taming,Sandholm00:eMediator,Nisan00:Bidding,Hoos01:Bidding}
can solve the communication overhead in some cases (when the bidder's
utility function is compressible).  However, they still require the
agents to completely determine and transmit their valuation functions
and as such do not solve all the issues.
So in practice, when the number of items for sale is even moderate, the
bidders will not bid on all bundles.
Instead, they may wastefully bid on bundles which they will not win, and
they may suffer reduced economic efficiency by failing to bid on bundles
they would have won.

Selective incremental preference elicitation by the auctioneer was
recently proposed to address these problems~\cite{Conen01:Preference},
but the idea was not evaluated.  We implemented the most promising
elicitation schemes from that paper, starting from a rigid
search-based scheme and continuing to a general flexible elicitation
framework.  We evaluated the previous schemes, and also developed a host of
new elicitation policies.
Our experiments show that elicitation
reduces revelation drastically, and that this benefit increases with
problem size.  We also provide theoretical results on elicitation
policies.  Finally, we introduce and evaluate a new query type that
takes the incremental nature of elicitation to a new level by allowing
agents to give approximate answers that are refined only on an
as-needed basis.


\mysection{Auction and elicitation setting}

We model the auction as having a single auctioneer selling a set $K$ of
items to $n$ bidder agents (let $k=|K|$).
Each agent $i$ has a {\em valuation function} $v_i: 2^K \mapsto \mathbb{R}$
that determines a finite private value $v_i(b)$ for each bundle $b
\subseteq K$.
We make the usual assumption that the agents have free disposal, that is,
adding items to an agent's bundle never makes the agent worse off because,
at worst, the agent can dispose of extra items for free.
Formally, $\forall b \subseteq K, b' \subseteq b$, $v_i(b) \geq v_i(b')$.
The techniques of the paper could also be used without free disposal,
although more elicitation would be required due to less \textit{a priori}
structure.

At the start of the auction, the auctioneer knows the items and the agents,
but has no information about the agents' value functions over the
bundles---except that the agents have free disposal.
The auction proceeds by having the auctioneer incrementally \emph{elicit}
value function information from the agents one query at a time until the
auctioneer has enough information to determine an optimal allocation of
items to agents.
Therefore, we also call the auctioneer the \emph{elicitor}.
An allocation is optimal if it maximizes social welfare $\sum_{i=1}^n
v_i(b_i)$, where $b_i$ is the bundle that agent $i$ receives in the
allocation.%
\footnote{
Social welfare can only be maximized meaningfully if bidders'
valuations can be compared to each other.  We make the usual
assumption that the valuations are measured in money (dollars) and thus
can be directly compared.  }
The goal of the elicitor is to determine an optimal allocation with as
little elicitation as possible.
A recent theoretical result shows that even with free disposal, in the
worst case, finding an (even only approximately) optimal allocation
requires exponential communication~\cite{Nisan02:Segal}.
Therefore, we will judge the techniques successful if they reduce
communication from full revelation by an asymptotic amount.

\mysection{Elicitor's inference and constraint network}

The elicitor, as we designed it, never asks a query whose answer could be
inferred from the answers to previous queries.
To support the storing and propagation of information received from the
agents, we have the elicitor store its information in a constraint
network.%
\footnote{This was included in the \emph{augmented order graph} of
Conen \& Sandholm~\cite{Conen01:Preference}.}
Specifically, the elicitor stores a graph for each agent.
In each graph, there is one node for each bundle $b$.
Each node is labeled by an interval $[\LB_i(b), \UB_i(b)]$.
The lower bound $\LB_i(b)$ is the highest lower bound the elicitor can
prove on the true $v_i(b)$ given the answers received to queries so far.
Analogously, $\UB_i(b)$ is the lowest upper bound.
We say a bound is {\em tight} when it is equal to the true value.

Each graph can also have directed edges.
A directed edge $(a,b)$ encodes the knowledge that the agent prefers bundle
$a$ over bundle $b$ (that is, $v_i(a) \geq v_i(b)$).
The elicitor may know this even without knowing $v_i(a)$ or $v_i(b)$.
An edge $(a,b)$ lets the elicitor infer that $\LB_i(a) \geq \LB_i(b)$,
which allows it to tighten the lower bound on $a$ and on any of $a$'s
ancestors in the graph.
Similarly, the elicitor can infer $\UB_i(a) \geq \UB_i(b)$, which allows it
to tighten the upper bound on $b$ and its descendants in the graph.

We define the relation $a \succeq b$ (read ``$a$ dominates $b$'') to be
true if we can prove that $v_i(a) \geq v_i(b)$.
This is the case either if $\LB_i(a) \geq \UB_i(b)$, or if there is a
directed path from $a$ to $b$ in the graph.
The free disposal assumption allows the elicitor to infer the following
dominance relations before the elicitation begins: $\forall b \subseteq K,
b' \subseteq b$, $b \succeq b'$.

%
Because the $\succeq$ relation is transitive, to encode the free
disposal constraints, we only need to add edges from each bundle $a$
to the bundles $b$ that include all but one item in $a$.  This allows
us to encode all the free disposal information in $k2^{k-1}$ edges per
agent\footnote{There are ${k \choose i}$ bundles with $i$ items.  A
bundle with $i$ items has $i$ outgoing edges (one for each item we
leave out).  Therefore, we have $\sum_{i=1}^k i {k \choose i} =
k2^{k-1}$ edges.}  rather than having to include in each graph one
edge for each of the $\half (3^k-1)$ dominance
relations.\footnote{Bundles with $i$ items have $2^{i-1}$ children
(every combination of $i$ items).  So, there are $\sum_{i=1}^k 2^{i-1}
{k \choose i} = \half (3^k-1)$ dominance relations.}

\mysection{Rank lattice based elicitation}

The elicitor can make use of non-cardinal rank information.
Let $b_i(r_i)$, $1 \leq r_i \leq 2^k$, be the bundle that agent $i$ has at
rank $r_i$.
In other words, $b_i(1)$ is the agent's most preferred bundle, $b_i(2)$ is
its second most preferred bundle, and so on until $b_i(2^k)$, which is the
empty bundle.

For example, consider two agents 1 and 2 bidding on two items $A$ and
$B$, and the following value functions:\\
\begin{tabular}{llll}
$v_1(AB)=8$,    & $v_1(A)=4$,   & $v_1(B)=3$, & $v_1(\emptyset)=0$ \\
$v_2(AB)=9$,    & $v_2(A)=1$,   & $v_2(B)=6$, & $v_2(\emptyset)=0$
\end{tabular} \\
So, agent 1 ranks $AB$ first, $A$ second, $B$ third, and the empty
bundle last.
Agent 2 ranks $AB$ first, $B$ second, $A$ third, and the
empty bundle last.

The elicitor uses a \emph{rank vector} $r=\rankvector{r_1, r_2, \ldots,
r_n}$ to represent allocating $b_i(r_i)$ to each agent $i$.
Not all rank vectors are feasible: the $b_i(r_i)$'s might overlap in items,
which would correspond to giving the same item to multiple agents.
For instance in the example above, rank vector $\rankvector{1,2}$
corresponds to allocating $AB$ to agent 1 and $B$ to agent 2, which is
infeasible.
Similarly, rank vector $\rankvector{2,2}$ allocates $A$ to agent 1 and $B$
to agent 2, which is a feasible allocation.
The value of a rank vector $r$ is $v(b(r)) = \sum_i v_i(b_i(r_i))$.
Rank vector $\rankvector{1,2}$ in our example has value $8+6=14$, while
$\rankvector{2,2}$ has value $4+6=10$.

The elicitor can put bounds on $v_i(b_i(r_i))$ using the constraint network
as before.
Even without knowing $b_i(r_i)$ (which bundle it is that agent $i$ values
$r_i$th), it knows that $v_i(b_i(r_i-1)) \leq v_i(b_i(r_i)) \leq
v_i(b_i(r_i+1))$.
Thus an upper bound on $v_i(b_i(r_i-1))$ is an upper bound on
$v_i(b_i(r_i))$, and a lower bound on $v_i(b_i(r_i+1))$ is a lower bound on
$v_i(b_i(r_i))$.
In our example, knowing only $b_1(1)=AB$ and $v_1(AB)=8$, the elicitor can
infer $v_1(b_1(2)) \leq 8$.

\begin{figure}[hbt]
\center{\epsfig{file=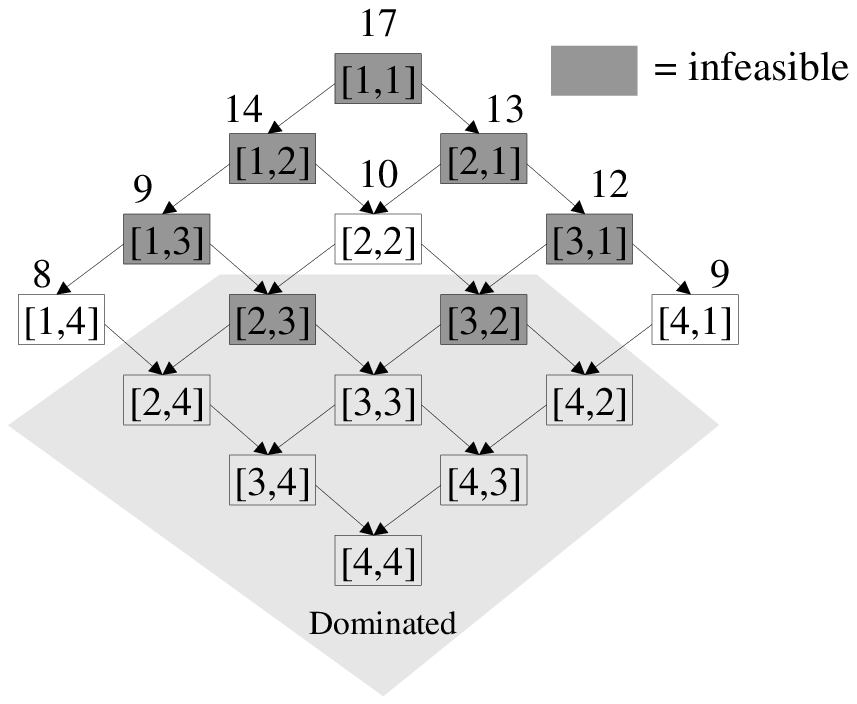,width=2.1in}}
\mycaption{\it Rank lattice corresponding to the example.  The gray nodes are
infeasible.  The shaded area is the set of nodes dominated by feasible
nodes.  The number above each node is the value of the node.  At the
outset, the auctioneer knows the structure of the lattice, but knows
neither the shadings nor the values of each node.}
\label{fi:rank-lattice}
\end{figure}

The set of all rank vectors defines a \emph{rank lattice}
(Figure~\ref{fi:rank-lattice}).
A key observation in the lattice is that the descendants of a node have
lower (or equal) value to the node.
%
%
%

Given the rank lattice, we can employ search algorithms to find an
optimal allocation.
In particular, by starting from the root and
searching in best-first order (always expanding the fringe node of
highest value), we are guaranteed that the first feasible node
that is reached is optimal.
%

{\algofontcheat\begin{algorithm}{FindOptimal}{}
    \textrm{FRINGE} \= \{ \langle 1,1,\ldots,1\rangle \} \\
    \begin{WHILE}{\textrm{FRINGE} \ne \emptyset}
        r = \CALL{FindBestNode}(\mathop{\textrm{FRINGE}}) \\
        \mathop{\textrm{FRINGE}} \= \mathop{\textrm{FRINGE}} - \{r\} \\
        \begin{IF}{\textrm{$r$ is feasible}}
            \RETURN r
        \end{IF} \\
        \begin{FOR}{\EACH r' \in \mathop{\textrm{children}}(r)}
          \begin{IF}{r' \notin \mathop{\textrm{FRINGE}}}
            \mathop{\textrm{FRINGE}} \= \mathop{\textrm{FRINGE}} \ \cup \ \{r'\}
          \end{IF}
        \end{FOR}
    \end{WHILE}
\end{algorithm}}\propcheat

Unlike in typical best-first search, algorithm \algofn{FindOptimal} does
not necessarily know which node of the fringe has highest value and thus
should be expanded next.
Determining this often requires more elicitation.
We implemented the following algorithm for doing this.
It corresponds to an elicitation policy where as long as we cannot prove
which node on the fringe is the best, we pick an arbitrary node and elicit
just enough information to determine its value.
%

{\algofontcheat\begin{algorithm}{FindBestNode}{\textrm{FRINGE}}
    S \= \textrm{FRINGE} \\
    \textrm{remove from $S$ all $r$ dominated by some $r'$ in $S$} \\
    \begin{IF}{\textrm{all $r\in S$ have the same value}}
        \RETURN \mathop{\textrm{arbitrary}} r \in S
    \end{IF} \\
    \textrm{choose $r \in S$ whose value we don't know exactly} \\
    \begin{FOR}{\EACH \mathop{\textrm{agent $i$}}}
        \begin{IF}{\textrm{elicitor does not know $b_i(r_i)$}}
            \textrm{ask agent $i$ what bundle it ranks $r_i$th}
        \end{IF}\\
        \begin{IF}{\textrm{elicitor does not know $v_i(b_i(r_i))$ exactly}}
            \textrm{ask agent $i$ for its valuation on bundle $b_i(r_i)$}
        \end{IF}
    \end{FOR} \\
    \algkey{goto} 2
\end{algorithm}}\propcheat

In some cases, \algofn{FindBestNode} can return a rank vector $r$ although
not all bundles $b_i(r_i)$ are known to the elicitor.
This can occur, for example, if the known valuations in the rank vector
already sum up to a large enough number.
In that case, checking the feasibility in step 5 of \algofn{FindOptimal}
requires eliciting the unknown bundles $b_i(r_i)$.

\mysection{Experimental setup}

While the idea and some algorithms for preference elicitation in
combinatorial auctions have been presented
previously~\cite{Conen01:Preference}, they have not been validated.
To evaluate the usefulness of the idea,
we conducted a host of experiments.
We present the results in the rest of the paper.
Each plot shows how many queries were needed to find an optimal allocation
and prove that it is optimal (that no other allocation is better).
In each plot, each point represents an average over 10 runs, where each run
is on a different problem instance (different draw of valuations for the
agents).
Each algorithm was tested on the same set of problem instances.

Because the evaluation is based on the amount of information asked rather
than real-time, we did not optimize our algorithm implementations for time
or space efficiency, but only for elicitation efficiency.
Generating all the plots in this paper took two days of computer time on a
1 GHz Pentium III.

Unfortunately, real data for combinatorial auctions are
not publicly available.\footnote{Furthermore, even if the data were
available, they would only have some bids, not the full valuation functions
of the agents (because not all agents bid on all bundles).}
Therefore, as in all of the other academic work on combinatorial auctions
so far, we used randomly generated data.
We first considered using existing
benchmark distributions.
However, the existing problem generators output instances with sparse bids,
that is, each agent bids on a relatively small number of bundles.
This is the case for the CATS suite of economically-motivated random
problem instances~\cite{Leyton00:Toward} as well as for the other prior
benchmarks~\cite{Sandholm02:Algorithm,Fujishima99:Taming,Andersson00:Integer,DeVries00:Combinatorial}.
In such cases, the communication is a non-issue, which undermines the
purpose of elicitation.
In addition, the instances generated by many of the earlier benchmarks do
not honor the free disposal constraints (because for an agent, the value of
a bundle can be less than that of a sub-bundle).

In many real settings, each bidder has a nonzero valuation for every
bundle.
For example in spectrum auctions, each bidder has positive value for every
bundle because each item is of positive value to every bidder (at least due
to renting and reselling possibilities).
In other settings, there may exist worthless items for some bidders.
Even in such cases, under the free disposal assumption, the bidders have
positive valuations for almost all bundles---except bundles that only
contain worthless items (because, at worst, they can throw away the extra
items in any bundle for free).

To capture these considerations, we developed a new benchmark problem
generator.
In each problem instance we generate, each bidder has a nonzero valuation
for almost every bundle, and all valuations honor free disposal.
Specifically, the generator assigns, for each agent in turn, integer
valuations using the following routine.
We impose an arbitrary maximum bid value $\mathrm{MAXBID} = 10^7$ in order
to avoid integer arithmetic overflow issues, while at the same time
allowing a wide range of values to be expressed.
Valuations generated with this routine exhibit both complementarity and
substitutability.

{\algofontcheat\begin{algorithm}{GenerateBids}{k}
    G \= \textrm{new constraint network} \\
    S \= 2^K \textrm{\hspace*{0.05in}(the set of all bundles)}\\
    \textrm{impose free disposal constraints on $G$} \\
    \UB(K) \= \mathrm{MAXBID} \\
    \begin{WHILE}{S \ne \emptyset}
        \textrm{pick $b$ uniformly at random from $S$} \\
        S \= S - {b} \\
        \textrm{pick $v(b)$ uniformly at random from $[\LB(b),\UB(b)]$} \\
        \textrm{propagate $\LB(b)=\UB(b)=v(b)$ through $G$}
    \end{WHILE}
\end{algorithm}}\propcheat

\mysection{Experiments on rank lattice based elicitation}

The first experiment evaluates the efficiency of rank lattice based
elicitation, see Figure~\ref{fi:rank}.
We plot the number of rank queries made (the number of value queries is
never greater because a value query is only ever asked after the
corresponding rank query).
For comparison, we plot the total number of value queries we could have
made: $n(2^k-1)$ (that is, for each agent, one query for each of the $2^k$
bundles except the empty bundle).
This corresponds to full revelation of each agent's valuation function.
Because this number grows exponentially in the number of items $k$, we use
a log scale on the vertical axis of the plot that shows performance as a
function of the number of items.
The other plot has a linear-scale vertical axis because full revelation
increases linearly in the number of agents $n$.

\makeplot{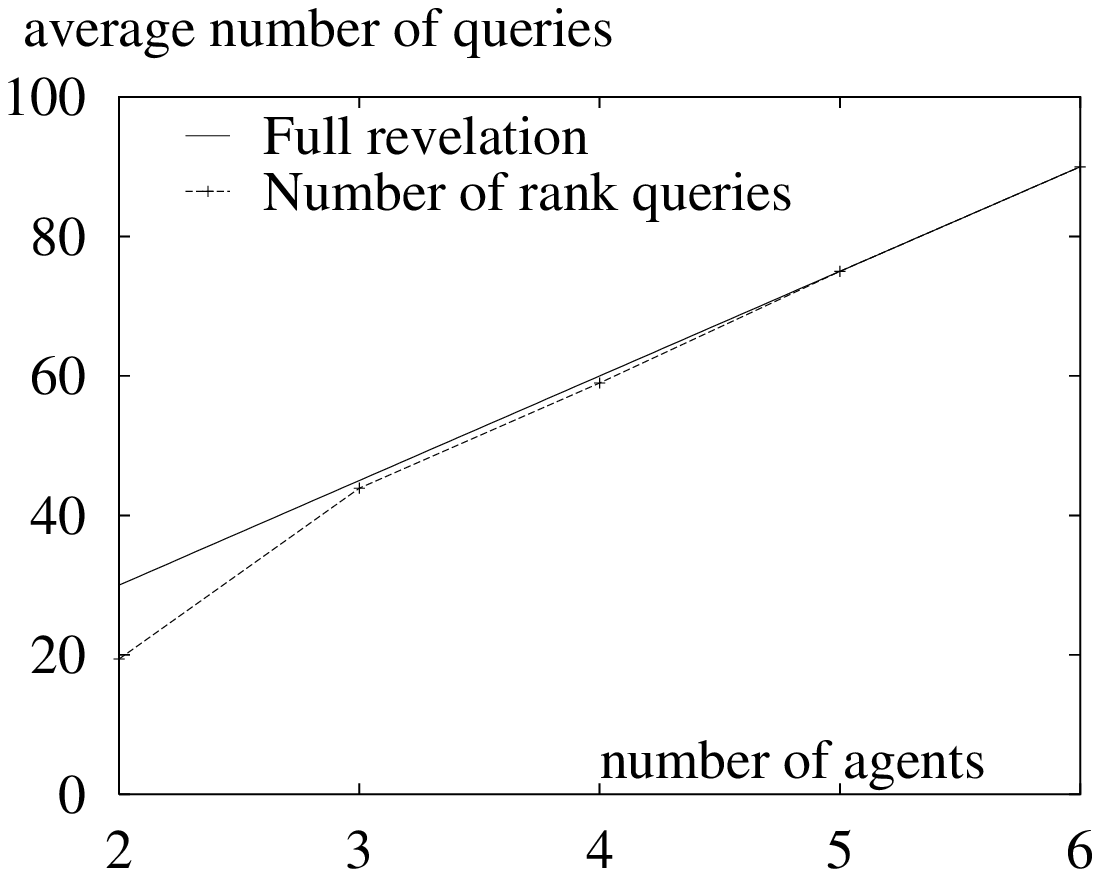}{Rank lattice based elicitation.  Left: 2 agents, varying number of items, log scale.  Right: 4 items, varying number of agents, linear scale.  All other graphs in the paper also follow this convention.}



Define the {\em elicitation ratio} to be the number of queries asked
divided by the number of queries asked in full revelation.
Figure~\ref{fi:rank} Left shows that as the number of items increases, the
elicitation ratio approaches zero, that is, only a vanishingly small
fraction of the possible queries are asked.

Figure~\ref{fi:rank} Right shows that as the number of agents $n$ grows,
the advantage from rank lattice based elicitation decreases.
This is not as important because even under full revelation, the number of
queries increases only linearly.
Nevertheless, this behavior might be explained by the observation that
while the size of the lattice grows exponentially in $n$, the number of
feasible nodes only grows polynomially.
Specifically, the total number of rank vectors is $(2^k)^n = 2^{nk}$ while
the number of feasible rank vectors is $n^k$ (each of the $k$ items can
independently go to any of the $n$ agents).
Therefore, as $n$ increases, this rank lattice based search procedure
encounters an increasing fraction of infeasible rank vectors before finally
finding an optimal allocation.

A very recent theoretical result shows that the algorithm here is as good
as any rank lattice based elicitation algorithm~\cite{Conen02:Partial}.
Specifically, this algorithm is a member of the \textbf{EBF} (efficient
best-first) family of algorithms, and the result proves that no algorithm
based on the rank lattice can guarantee asking fewer queries than an
\textbf{EBF} algorithm over all problem instances (unless it sacrifices
economic efficiency).

\mysection{General elicitation framework}

Given that no rank lattice based algorithm can do better than the one
outlined above, we now move to a more general elicitation framework.
As we will show, this allows us to develop algorithms that ask
significantly fewer queries.

The framework allows a richer set of query types (to accommodate for
different settings where answering some types of queries is easier than
answering other types); allows more flexible ordering of the queries at run
time; and never considers infeasible solutions.
We could implement rank queries in this framework, but did not do so in
this work, because rank queries are somewhat unrealistic: to answer them
would likely require the bidder to evaluate and sort its entire valuation
function.

The general algorithm template is a slightly modified version of that
of Conen \& Sandholm~\cite{Conen01:Preference}:
{\algofontcheat\begin{algorithm}{Solve}{}
        C \= \CALL{InitialCandidates}(n,k) \\
        \begin{WHILE}{\algkey{not} \CALL{Done}(C)}
            q \= \CALL{SelectQuery}(C) \\
            \CALL{AskQuery}(q) \\
            C \= \CALL{Prune}(C)
        \end{WHILE}
\end{algorithm}}\propcheat

Here, $C$ is a set of candidates, where a \emph{candidate} is a vector $c =
\langle c_1, c_2, \ldots, c_n\rangle$ of bundles where the bundles contain
no items in common.
Unlike with rank vectors, all candidates are feasible.
The value of a candidate is $v(c) = \sum_i v_i(c_i)$; $\UB(c)=\sum_i
\UB_i(c_i)$ is an upper bound, and $\LB(c)=\sum_i \LB_i(c_i)$ a lower
bound.
A candidate $c$ dominates another candidate $c'$ if the elicitor can prove
that the value of $c$ is at least as high as that of $c'$.%
\footnote{This is the case if $\LB(c) \geq \UB(c')$.  Even if not, the
elicitor can use the edges in the graph.  If there is a subset of the
agents $I$ such that $\forall i \in I$, $c_i \succeq c_i'$, and that for
the remaining agents, $\sum_{j \notin I} \LB_j(c_j) \geq \sum_{j \notin I}
\UB_j(c_j)$, then this also constitutes a proof that $c$ has value at least
as high as $c'$.}

\algofn{InitialCandidates} generates the set of all candidates, which is
the set of all $n^k$ allocations of the $k$ items to the $n$ agents (some
agents might get no items).
In our experiments, the candidate set is represented explicitly.
To scale the implementation to large $k$ and $n$ would require representing
it more intelligently in an implicit way.

\algofn{Prune} removes, one candidate at a time, each candidate that is
dominated by a remaining candidate.
This may eliminate some optimal allocations, but it will never eliminate
all optimal allocations---one will always remain.
If strict domination were to be used as the criterion, then \algofn{Solve}
would find \emph{all} optimal allocations, at the cost of requiring more
elicitation.



\algofn{Done} returns true if $C$ is a set of candidates, each of which is
provably optimal.
This is the case either if $C$ has only one element, or if all candidates
in $C$ have known value (that is, $\forall c \in C, \UB(c)=\LB(c)$).
Because the algorithm has just pruned, it knows that if all candidates have
known value, then they have equal value.

\algofn{SelectQuery} selects the next query to be asked.
This function can
be instantiated in different ways to implement different elicitation
policies, as we will show.

\algofn{AskQuery} takes a query, asks the corresponding agent for the
information, and appropriately updates the constraint network.
The details of updating the network are discussed in conjunction with each
query type below.

\mysubsection{Value queries}

The most basic query asks an agent $i$ to reveal $v_i(c_i)$ exactly.
We call such queries {\em value queries}.
Upon receiving the answer, \algofn{AskQuery} sets
$\LB_i(c_i)=\UB_i(c_i)=v_i(c_i)$ and propagates the new bounds upstream and
downstream through the constraint network as described earlier.

Any policy that asks only value queries relies on there being edges in the
constraint network, for instance due to free disposal.
Otherwise, it needs to ask every query: any value the elicitor has not
asked for might be infinite.

\mysubsubsection{Random elicitation policy}

The first policy we investigate simply asks random value queries.
In the beginning, we generate the set of all $n(2^k-1)$ value queries.
Whenever it is time to ask a query, the policy chooses a random query
from the set, ignoring those it has already asked or for which the
value can already be inferred.

We can actually show that if any policy saves elicitation, then this
policy also saves elicitation:%

\propcheat\begin{prop}
\label{prop:random_elicitor}
Let $Q= n (2^{k}-1)$ be the total number of queries, and let $\qmin$
be the number of queries asked by an optimal elicitation policy.  For
any given problem instance, the expected number of queries that the
random elicitation policy asks is at most $\frac{\qmin}{\qmin +1}
(Q +1)$.
\end{prop}\propcheat

\begin{proof}
Assume pessimistically that a query is either required to prove the
optimal allocation or useless.
%
%
Under this assumption, the analysis reduces to the following problem.
We have $r$ red ``necessary'' balls and $b$ blue ``useless'' balls in
a bag.  We then randomly draw one ball at a time without
replacement.  The question is how many balls we expect to draw before
all red balls have been drawn.  Let $e(r,b)$ be this number.  The base
case is $e(0,b)=0$, because there are no red balls to draw.  In the
general case, we pick one ball from the bag.  With probability
$r/(r+b)$, it is red, so the bag now has $r-1$ red balls and $b$ blue
balls.  Similarly, with probability $b/(r+b)$, it is blue, so the bag
now has $r$ red balls and $b-1$ blue balls.  Therefore, $e(r,b) = 1 +
\frac{r}{r+b} e(r-1,b) + \frac{b}{r+b} e(r,b-1)$.  It is easy to
verify that $e(r,b) = \frac{r}{r+1}(b+r+1)$ solves this recurrence.
In the elicitation setting, $r=\qmin$ and $r+b =Q$.  The result
follows.
\end{proof}

The upper bound given in the above proposition only guarantees
relatively minor savings in elicitation (especially because $\qmin$
increases when the number of agents and items increases).  This could
be due to either the bound being loose, or due to this elicitation
policy being poor, or both.  The experiment in Figure~\ref{fi:vrand}
shows that this elicitation policy is poor---even in the average
case. The policy asks almost all of the queries.


\makeplot{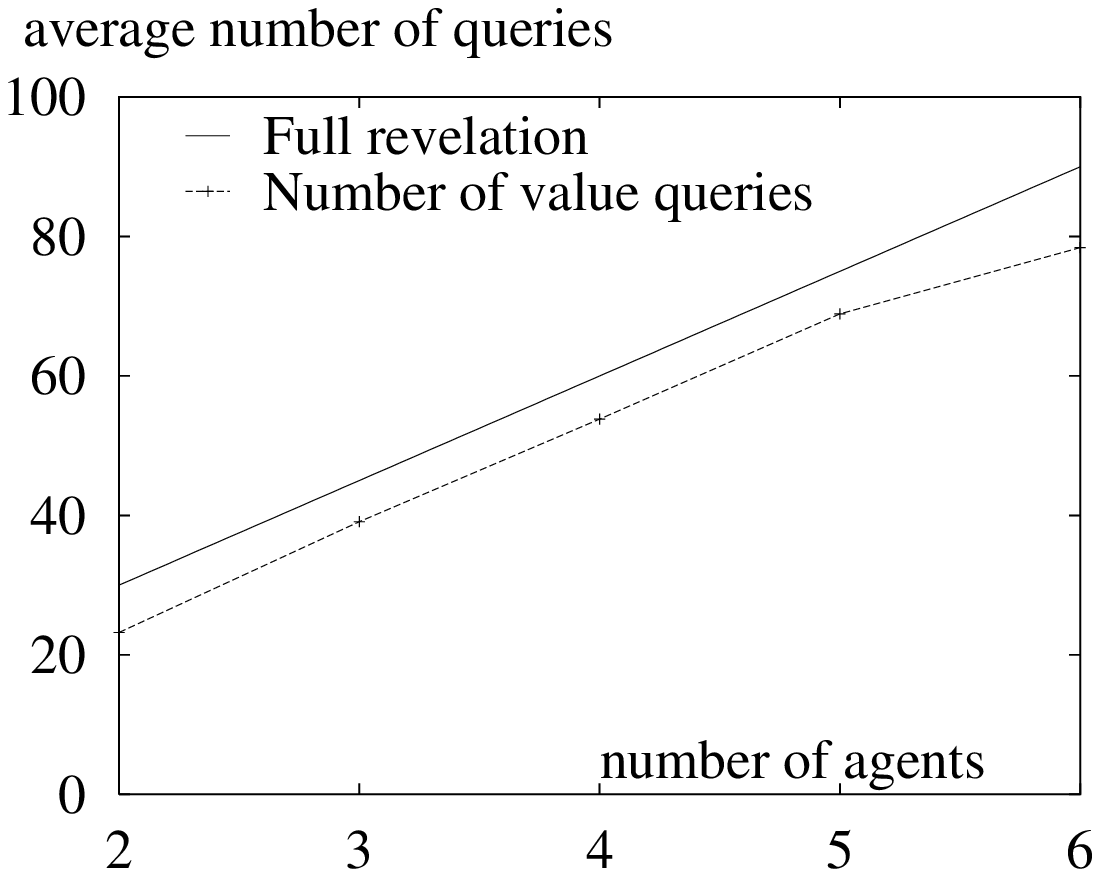}{Random elicitation policy.}

\mysubsubsection{Random allocatable bundle elicitation policy}

Essentially, the random elicitation policy asks many queries which, as
it turns out, are not useful.  We will now present a useful
restriction on the set of queries from which the elicitation policy
should choose.  The key observation is that the elicitor might already
know that a bundle is not going to be allocated to a particular
bidder---even before the elicitor knows the bidder's valuation for the
bundle.  This can occur if the elicitor knows that it cannot obtain
enough value from the other bidders for the items not in that bundle.
On the other hand, if the elicitor cannot (yet) determine this, then
the bundle-agent pair is called an {\em allocatable}.

\propcheat\begin{defin}
A bundle-agent pair $(b,i)$ is {\em allocatable} if there exists a
remaining candidate allocation $c \in C$ such that $c_i =b$.  In some
places, the reference to the agent is obvious from the context, so we
sometimes talk about allocatable bundles $b$ rather than $(b,i)$.
\end{defin}\propcheat

Now we can refine our random elicitation policy to ask queries on
allocatable $(b,i)$ only (and queries that have already been asked or whose
answer can be inferred are again never asked).

This restriction is intuitively appealing, and we can characterize
cases where it cannot hurt. We define the notation $\alloccase{x,y}$
to mean that revealing the value of a non-allocatable pair $(b,i)$
would raise the lower bound on $x$ allocatable super-bundles of $b$
(that is, there are $x$ allocatable pairs $(b',i)$ such that
$b'\succeq b$), and lower the upper bound on $y$ allocatable
sub-bundles of $b$.
To affect a lower bound, $v_i(b)$ must be strictly greater
than the currently-proven lower bound on any of the $x$ super-bundles.
Similarly, $v_i(b)$ must be strictly less than the currently-proven
upper bound on the $y$ sub-bundles.

Given this notation, we can examine the cases where eliciting a
non-allocatable $(b,i)$ is no more useful than eliciting some
allocatable $(b',i)$.  Because the elicitor does not know $v_i(b)$, it
cannot know what case actually applies.

\propcheat\begin{prop}
\label{prop:good_cases}
No matter what value queries the elicitor has asked so far, querying a
non-allocatable $(b,i)$ in case $\alloccase{x,y}$ with $x+y < 2$ cannot
help the unrestricted random elicitation policy ask fewer queries than the
restricted random elicitation policy.
\end{prop}\propcheat

\begin{proof}
We analyze each case separately.

\textbf{Case $\alloccase{0,0}$}:
In this case, $b$ has no allocatable sub- or super-bundles, so
obtaining values for those bundles cannot be useful.  So, if $b$
itself is not allocatable, eliciting a value for $b$ cannot be useful.
%
%
(Here, ``useful'' means that it helps prove that an allocation is
optimal or it helps prove that an allocation is not optimal.)

\textbf{Cases $\alloccase{0,1}$ and $\alloccase{1,0}$}:
The two cases are symmetric; we assume $\alloccase{0,1}$ here.  Let
the single allocatable sub-bundle of $b$ be called $b'$.  By eliciting
$v_i(b)$, the upper bound on $v_i(b')$ will be tightened.  However,
eliciting $v_i(b')$ would reveal an upper bound on $b'$ that is at
least as tight.  So eliciting the allocatable bundle $b'$ would have
been no worse.  In fact, it might have been strictly better: with the
same number of queries, eliciting $v_i(b')$ also reveals a lower bound
on $b'$ and any super-bundle of $b'$ (in particular, any allocatable
super-bundle).
\end{proof}

While the idea of restricting the queries to allocatable bundles is
intuitively appealing and can never hurt in the cases above, there are
cases where this restriction forces the elicitor to ask a larger
number of queries:

\propcheat\begin{prop}
\label{prop:exists_bad_case}
Querying a non-allocatable $(b,i)$ in case $\alloccase{x,y}$ with $x+y \geq
2$ may help the unrestricted random elicitation policy ask fewer queries
than the restricted random elicitation policy.
\end{prop}\propcheat
\begin{proof}
Assume there are two bidders.
Further assume that given the information elicited previously, there are
only three allocations that could be optimal ($|C|=3$).
One allocation involves giving bidder 1 the items in bundle $b'$, and
the other items to bidder 2, and bidder 2 places a value of 50 on
those items.  Similarly, the second allocation gives bidder 1 the
items in bundle $b''$, and the other items to bidder 2, who again
places a value of 50 on those items.  Bundle $b'$ is neither a
super-bundle nor a sub-bundle of $b''$.  Given the information
elicited so far, the elicitor knows $\UB_i(b')=\UB_i(b'')=100$.  A
third allocation gives bidder 2 all the items, and bidder 2 places a
value of 100 on this outcome; bidder 1 gets no items.  Finally, assume
the true value agent 1 has on bundle $b$ is 40 ($v_1(b)=40$),
and that $b$ is not allocatable for agent 1.  This means that $(b,1)$ is in
case $\alloccase{0,2}$.

\begin{table}[hbt]
\begin{tabular}{lll}
$v_1(b')=[0,100]$   & $v_2(K-b')=50$   & sum: $[0,150]$\\
$v_1(b'')=[0,100]$  & $v_2(K-b'')=50$  & sum: $[0,150]$\\
$v_1(\emptyset)=0$  & $v_2(K)=100$     & sum: $100$\\
\end{tabular}
\mycaption{\it Example of case $\alloccase{0,x}$ where revealing a
non-allocatable bundle $b$ is better than revealing any allocatable bundle.
By eliciting $v_1(b)=40$, the elicitor learns that the first two candidate
allocations have a value of at most 90, and can therefore eliminate them.}
\end{table}

In this situation, by eliciting just $v_1(b)$, the elicitor would prove
that the third allocation (giving bidder 2 all the items) is optimal.
Restricted to revealing allocatable bundles $b'$ and $b''$ but not
$b$, the elicitor will instead need to use two queries instead of just one.
%
%
\end{proof}

Proposition~\ref{prop:exists_bad_case} does not mean that all cases
$\alloccase{x,y}$ with $x+y \geq 2$ are bad.
Indeed, it could be that eliciting the non-allocatable $(b,i)$ gives
insufficiently tight bounds on the allocatable bundles it affects, and
therefore the allocatable bundles need to be elicited anyway.
An open problem is whether, in the case of an oracle that chose the
best bundle to elicit every time, the bad cases would ever happen.
%

The case-by-case analysis of Propositions~\ref{prop:good_cases}
and~\ref{prop:exists_bad_case} indicates that restricting the elicitation
policy to choosing only allocatable bundles will often help, but may
sometimes also cause harm.  However, the harm is limited, as we
now show.

\propcheat\begin{prop}
\label{prop:bad_case_bound}
Any bad case $\alloccase{x,y}$ with $x+y \geq 2$ causes the random query
policy that restricts itself to allocatable queries to ask at most
twice as many queries (in expectation) as the unrestricted random policy.
This bound is tight.
\end{prop}\propcheat

\begin{proof}
Assume that to prove an optimal allocation, it is necessary either to
reveal the single non-allocatable bundle, or \emph{all} of the $x+y$
allocatable bundles.  If fewer than all $x+y$ bundles are needed, or if
there is more than one subset of the $x+y$ that is sufficient, this only
reduces the advantage of being allowed to ask the bad-case query.

In the restricted policy, which cannot ask the bad-case query, the elicitor
will ask $x+y$ queries.  In the unrestricted policy, the elicitor's task
corresponds to removing balls one at a time from a bag that has 1 red ball
and $b=x+y$ blue balls until either the red ball has been removed, or all
the blue balls have been removed.  Let $e(b)$ be the number of balls we
expect to pick until we are done.  We pick a red ball with probability
$1/(b+1)$ and are done immediately.  Otherwise, we pick a blue ball
with probability $b/(b+1)$.  Therefore, $e(b)=\frac{1}{b+1} +
\frac{b}{b+1}(1+e(b-1))$.  If only one blue ball is left, whether we pick
the red ball or the blue ball, we are done, so $e(1)=1$.  It can be
verified that $e(b)=\frac{b(3+b)}{2(1+b)}$ solves the recurrence relation.
Therefore, the ratio of queries asked in the restricted policy to queries
asked in the unrestricted policy is $(x+y)/e(x+y) \leq 2$.
\end{proof}

Summarizing, restricting value elicitation to allocatable bundles either helps,
does not hurt, or at worst only causes the elicitor to ask (in expectation)
twice as many queries.
We ran experiments (Figure~\ref{fi:vrand_unif}) to determine whether the
restriction helps in practice.  The results are clear: at $k=10$, the
elicitation ratio is $17\%$.  That is, the random elicitation policy
restricted to eliciting only allocatable $(b,i)$ avoids the vast majority
of the elicitation needed in full revelation or in the unrestricted random
elicitation policy.  Most importantly, as the number of items increases,
the elicitation ratio continues to decrease (unlike with the random
elicitation policy without the restriction).  Also, unlike with rank
lattice based elicitation, as the number of agents increases, the
elicitation ratio stays constant or may even decrease.

\makeplot{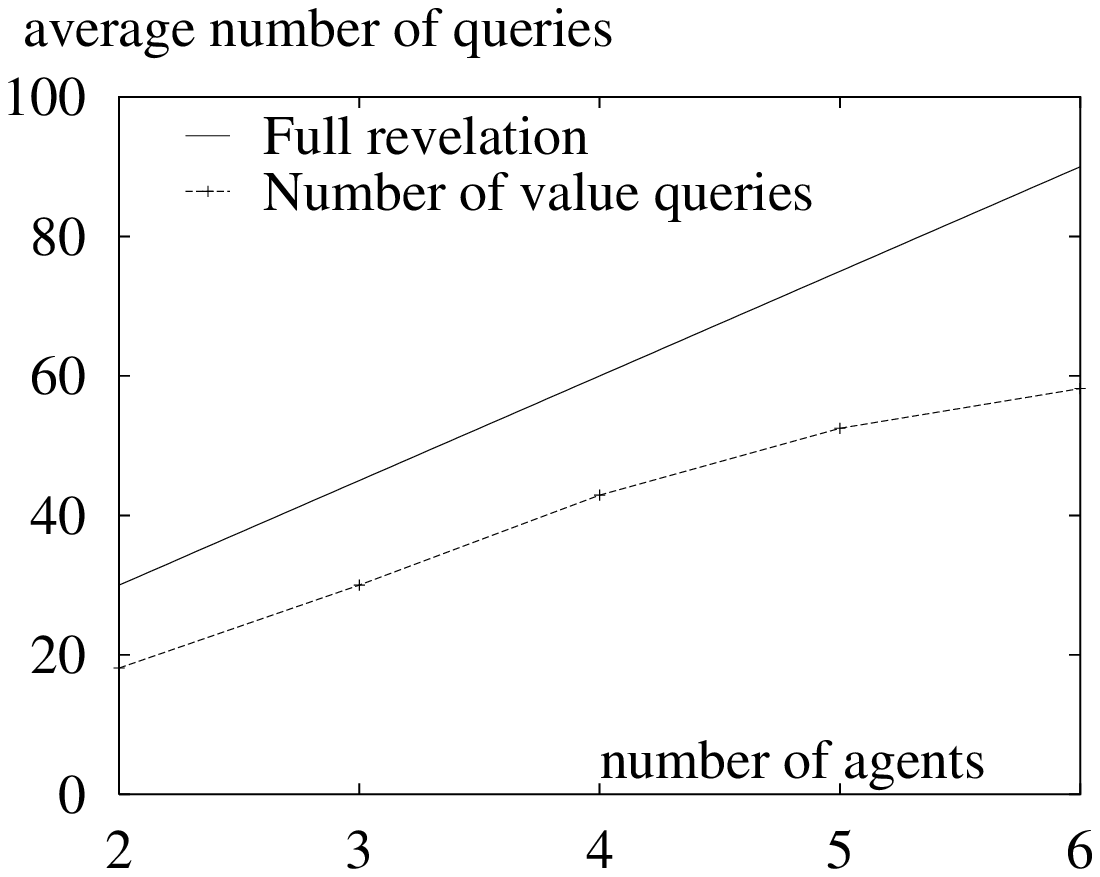}{Random allocatable-only elicitation policy.}

%
%

This policy is simpler than the value-query policy previously
proposed by Conen \& Sandholm~\cite{Conen01:Preference}.
That policy counted the number of remaining candidates in which a given
bundle $b$ is allocated to agent $i$, and elicited the value of the $(b,i)$
pair with the highest count.  We ran the same experiment with that policy.
Interestingly, that policy does far less well:  depending on the
tie-breaking scheme, it asked exactly all the queries (when breaking ties
in favor of the smaller bundle, arbitrarily choosing among equal-size
bundles), or it converged to about half the queries (when breaking ties in
favor of the larger bundle).  A randomized tie-breaking scheme did slightly
worse than breaking ties in favor of larger bundles.

\mysubsubsection{The grand bundle is (almost) always revealed}

Intuitively it is appealing to elicit from every agent the value for
the grand bundle because that sets an upper bound on all bundle-agent
pairs (via the free disposal assumption).  In this section we analyze
whether this indeed is a good idea.
%

\propcheat\begin{prop}
\label{prop:grand_bundle}
In order to determine the optimal allocation, any elicitation policy must
prove an upper bound on $v_i(K)$ for every $i$ to which $K$ is not
allocated.
\end{prop}\propcheat
\begin{proof}
The lower bound on the optimal allocation is finite (say, $L$) because
we require each bundle to have non-negative and finite value for every
bidder.  Therefore, unless the auctioneer provides an upper bound on
$v_i(K)$, the possibility is open that allocating $K$ to $i$ is worth
more than $L$.
Because allocating $K$ to $i$ possibly has value greater
than implementing the allocation that is, in fact, optimal, the elicitation
policy cannot terminate.
\end{proof}

In particular, using value queries only (and with no extra structure beyond
free disposal), the only way the auctioneer can establish an upper bound on
$v_i(K)$ is by eliciting the value.

\propcheat\begin{theorem}
Assume there are at least 2 bidders.  There is a policy (possibly
requiring an oracle for choosing the queries) using value queries that asks
$v_i(K)$ for every $i$ and that asks the fewest possible questions.
\end{theorem}\propcheat

\begin{proof}
If the optimal allocation involves allocating items to at least two
bidders, then we are not allocating the full bundle to any agent, so by the
proposition above, the auctioneer must elicit $v_i(K)$ for every $i$.

Otherwise, the optimal allocation involves allocating all items to a single
agent $i$.  For all $j \neq i$, the proposition applies and therefore the
auctioneer must elicit $v_j(K)$.  What is left to prove is that the
auctioneer is at least as well off also eliciting $v_i(K)$.

If all other bidders have zero value on all sub-bundles (that is, the full
bundle has positive value to them, but anything less has zero value), then
the auctioneer need only place a lower bound on $v_i(K)$ that is higher
than any other bidder's, which it can do by eliciting $v_i(K)$.

If any other bidder $j$ has nonzero value on some bundle $K-b$, the
auctioneer needs to prove that $v_j(K-b)+v_i(b) \leq v_i(K)$.  In other
words, the auctioneer needs to provide a lower bound on $v_i(K)$ that
is sufficiently greater than the upper bound on $v_i(b)$.  Using value
queries and assuming free disposal, the auctioneer can prove an upper
bound on $v_i(b)$ by eliciting $b$ or any super-bundle $b'$.
Similarly, it can prove a lower bound on $K$ by eliciting $K$ or any
sub-bundle.  However, it cannot prove sufficiently tight bounds to
separate the optimal allocation (of $K$ to $i$) from the suboptimal
one by eliciting a single bundle: by eliciting a single bundle, it
would prove $\UB_i(b)=\LB_i(K)$.  Given that it must elicit two
bundles, it may as well elicit $v_i(K)$ to provide the lower bound on
that value.

The argument in the paragraph above generalizes to settings where
there are many bidders $j$ who have nonzero value for several bundles
$K-b$.  The auctioneer must prove a sufficiently tight lower bound on
$K$, and sufficiently tight upper bounds on each of the bundle-agent
pairs $(b,i)$.  Since not all the elicitations that support
sufficiently tight upper bounds on $b$'s can also support a
sufficiently tight lower bound on $K$, the auctioneer must elicit at
least one other bundle to support that lower bound.  There may be more
than one choice for this; however, one possible choice is simply to
elicit $v_i(K)$.
\end{proof}

If $n=1$, the statement does not hold because without revealing anything,
we already know by free disposal that giving the bidder all of $K$
is an optimal allocation.

\mysubsection{Order queries}

In some applications, agents might not know the values of bundles, and
might need to expend a lot of effort to determine
them~\cite{Sand93,Larson01:Deliberation}, but might easily be able to
see that one bundle is preferable over another.  In such settings, it
would be sensible for the elicitor to ask {\em order queries}, that
is, ask an agent $i$ to order two given bundles $c_i$ and $c_i'$ (to
say which of the two it prefers).  The agent will answer $c_i
\succeq c_i'$ or $c_i' \succeq c_i$ or both.
\algofn{AskQuery} will then create new edges in the constraint network to
represent these new dominates relations.
By asking only order queries, the elicitor cannot compare the
valuations of one agent against those of another, so it cannot
determine a social welfare maximizing allocation.  However, order
queries can be helpful when interleaved with other types of queries.

\mysubsection{Using value and order queries}

We developed an elicitation policy that uses both value and order
queries.  It mixes them in a straightforward way, simply alternating
between the two, starting with an order query.  Whenever an order
query is to be asked, the elicitor picks an arbitrary pair $(c,c')$ of
remaining candidates that cannot be compared due to lack of
information, chooses an agent $i$ whose ranking of $c_i$ and $c_i'$ is
unknown, and asks that agent to order bundles $c_i$ and $c_i'$. (This
is the policy for choosing order queries that was proposed
by Conen \& Sandholm~\cite{Conen01:Preference}.)  Whenever a value query is
to be asked, the query is chosen using the policy described in the value
query section above.

To evaluate the mixed policy, we need a way of comparing the cost of an
order query to the cost of a value query.  The plots in this section
correspond to a cost model where an order query costs 10\% of the cost of a
value query.

Figure~\ref{fi:vo1} shows that the amount of elicitation grows
linearly with the number of agents.  Also, as the number of items
increases, the cost of the queries is a vanishing fraction of the cost
of full elicitation.

\makeplot{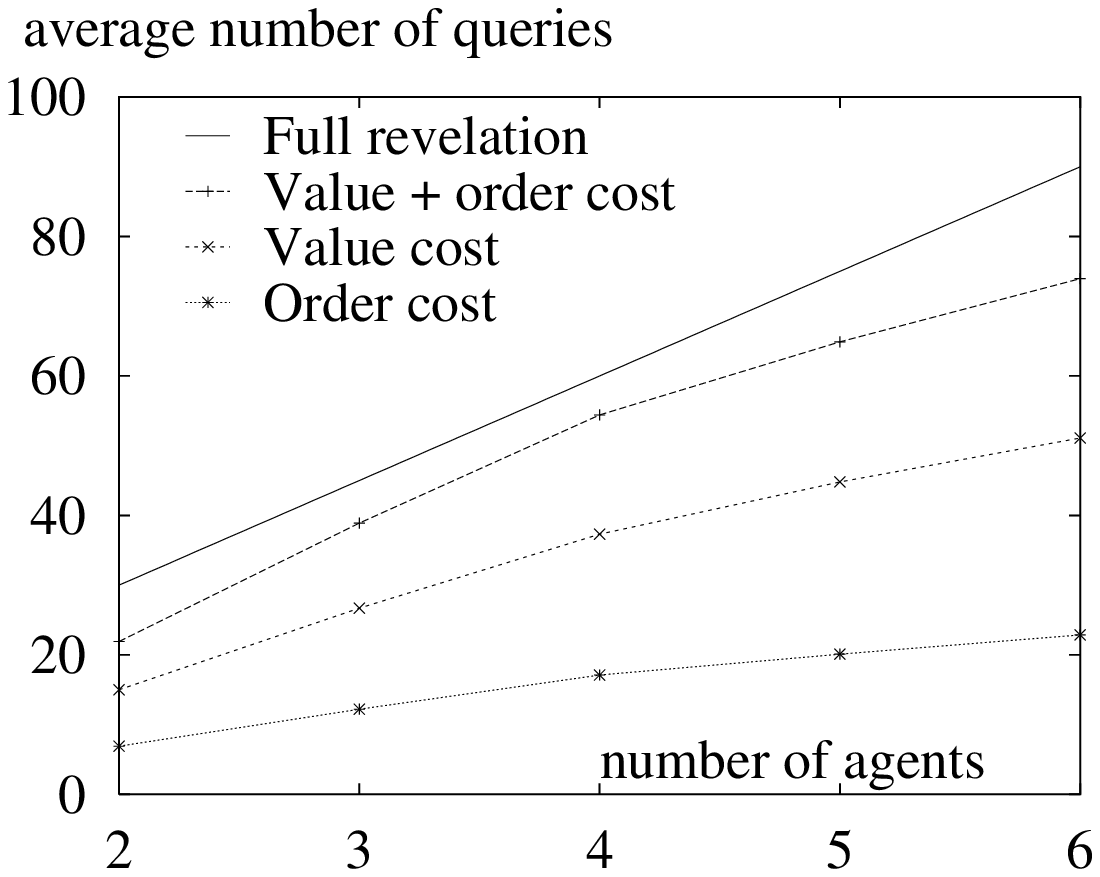}{Elicitation using value and order queries.}


The policy saves elicitation cost compared to the policy that only uses
value queries.  For example, at 2 agents and 10 items, its elicitation cost
averages 324 while the elicitation cost of the value-only policy averages
361.
%
%
As the relative cost of an order query is decreases, the benefit of
interleaving value queries with order queries increases.  Also, if order
queries are inexpensive, the policy should probably ask more than one order
query per value query.


While this mixed policy appears to provide only a modest benefit over
using value queries only, its advantage is that it does not depend as
critically on free disposal. Without free disposal, the policy that
uses value queries only would have to elicit all values.  The order
queries in the mixed policy, on the other hand, can create useful
edges in the constraint network which the elicitor can use to
prune candidates.

\mysubsection{Bound-approximation queries}

In many settings, the bidders can roughly estimate valuations easily, but
the more accurate the estimate, the more costly it is to determine.
In this sense, the bidders determine their valuations using anytime
algorithms~\cite{Larson01:Deliberation,Larson01:Computationally}.
For this reason, we introduce a new query type: a {\em bound-approximation
query}.
In such a query, the elicitor asks an agent $i$ to tighten the agent's
upper bound $\UB_i(b)$ (or lower bound $\LB_i(b)$) on the value of a given
bundle $b$.
This query type leads to more incremental elicitation in that queries are
not answered with exact information, and the information is refined
incrementally on an as-needed basis.

The elicitor can provide a hint $t$ to the agent as to how much
additional time the agent should devote to tightening the bound in the
query.  Smaller values of the hint $t$ make elicitation more
incremental, but cause additional communication overhead and
computation by the elicitor.  Therefore, the hint can be tailored to
the setting, depending on the relative costs of communication, bundle
evaluation by the bidders, and computation by the elicitor.  The
hint could also be adjusted at run-time, but in the experiments below, we
use a fixed hint $t=0.2$.


To evaluate this elicitation method, we need a model on how
the agents' computation refines the bounds.  We designed the details of
our elicitation policy motivated by the following specific scenario,
although the elicitation policy can be used generally.  Let each agent
have two anytime algorithms which it can run to discover its value of
any given bundle: one gives a lower bound, the other gives an upper
bound.
%
%
Spending time $d$, $0 \leq d \leq 1$ will yield a lower bound
$v_i(b)\sqrt{d}$ or an upper bound
$(2-\sqrt{d})v_i(b)$.
This means that there are diminishing
returns to computation, as is the case with most anytime
algorithms.\footnote{The square root is arbitrary, but captures the case of
diminishing returns to additional computation.  Running experiments with
$d$ in place of $\sqrt{d}$ did not significantly change the results.}
Finally, we assume that the algorithms can be restarted from the best
solution found so far with no penalty: having spent $d$ time tightening a
bound, we can get the bound we would have gotten spending $d'>d$ by only
spending an additional time $d'-d$.
%

The model of agents' computation cost here opens the possibility to cheat
in the evaluation of the elicitor.  As the model is stated, the elicitor
could ask an agent to spend $t$ time each on the upper and lower bound.
Based on the answers, the elicitor would know the exact value (it would be
in the middle between the lower and upper bound).  To check that our
results do not inadvertently depend on such specifics of the agents'
computation model, we ran experiments using an asymmetric cost function
(linear for lower bounds, square root for upper bounds).  This did not
appreciably change the results.

%
%
%
Using arbitrarily picked bound-approximation queries as the
elicitation policy would work, but the more sophisticated elicitation
policy that we developed chooses the query that maximizes the
%
%
expected benefit.
%
%
This is the amount by which we expect the upper and lower bounds on
bundle-agent pairs to be tightened when we propagate the new bound
that the queried agent will return (only counting bundle-agent pairs
that are included in the set of remaining candidates).
To compute the expected benefit, the elicitor assumes that $v_i(b)$ is
drawn uniformly at random in $[\LB_i(b),\UB_i(b)]$.
%
%
To estimate the expected change in bounds, the elicitor samples 10 equally
spaced
values $\hat{v_i}(b)$ in that interval.  For each
value, the elicitor computes (using the cost model described in the
previous paragraph) what bound $z$ it would receive if the agent spent
additional time $t$ working on that bound and the true value were
$\hat{v_i}(b)$.  Finally, the elicitor observes by how much the values
in the constraint network would change if the elicitor were to
propagate $z$ through the network (only bundle-agent pairs that are
included in the set of remaining candidates are counted).\footnote{A
minor detail comes in estimating the worth of reducing an upper bound
from $\infty$.  We dodge this question by initially asking each agent
for an upper bound on the grand bundle---which is almost always required
as shown in Proposition~\ref{prop:grand_bundle}.  By free
disposal, that is also an upper bound on all other bundles.}

We evaluated bound-approximation queries using the elicitation policy
and agents' computation model described above.  Figure~\ref{fi:vba}
shows that as the number of items increases, only a vanishingly small
fraction of the overall computation cost is actually incurred because
the optimal allocation is determined while querying only very
approximate valuations on most bundle-agent pairs.  The method also
maintains its benefit as the number of agents increases.
%

\makeplot{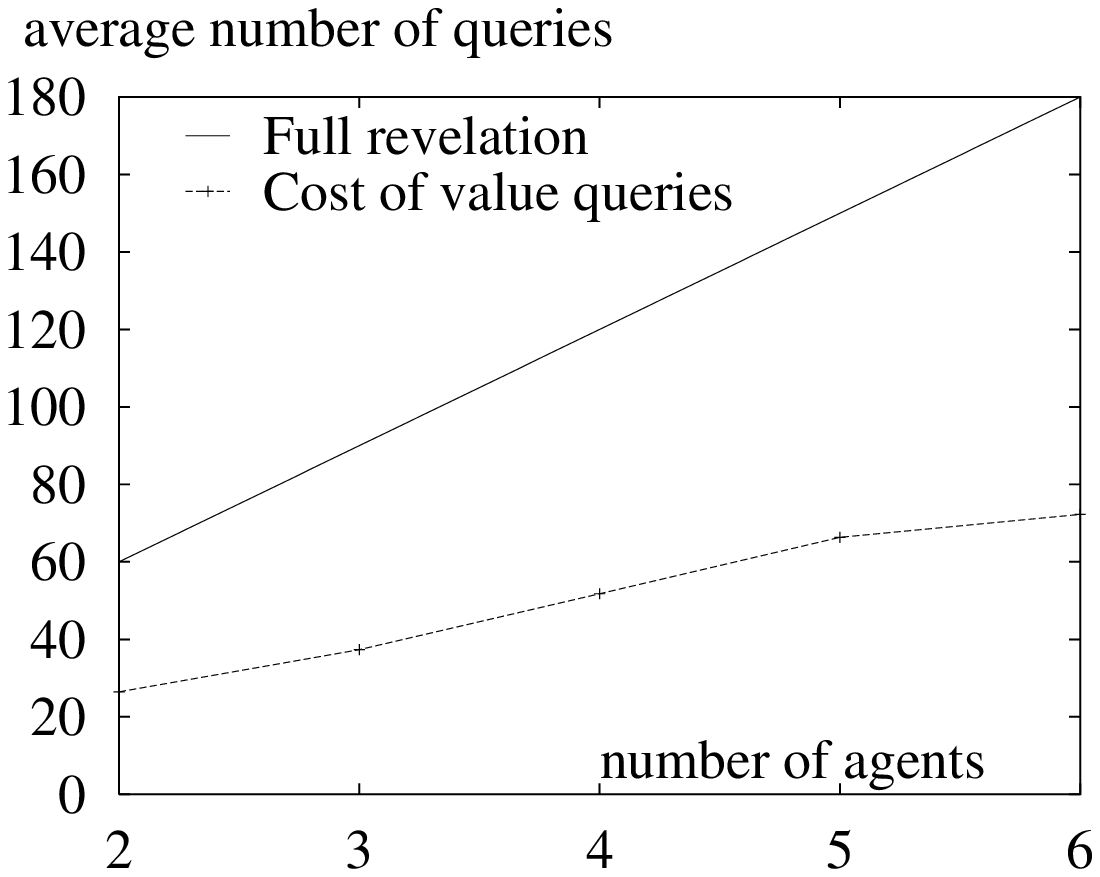}{Elicitation using bound-approximation queries.  Because it costs $1$ to get a tight upper or lower bound, it costs $2$
to make both bounds be tight.  Thus the worst-case line in these plots
is $2n(2^k-1)$.}

\mysubsubsection{Using bound-approximation and order queries}

As in the policy that mixed value and order queries, we can alternate
between bound-approximation queries and order queries.
Figure~\ref{fi:vo1vba} presents the results when bound-approximation
queries are charged as in the previous section, and order queries are
charged $\frac{1}{10}$.
As the number of agents increases, only a vanishingly small fraction
of the cost of full revelation ends up being paid.  The method also
maintains its benefit as the number of agents grows.
%
%
The policy saves elicitation cost compared to the policy that only uses
value queries.  For example, at 2 agents and 8 items, its elicitation cost
averages 172 while the elicitation cost of the policy that only uses
bound-approximation queries averages 230.  Furthermore, as the relative
cost of order queries is lowered, the mixed method becomes increasingly
superior to using bound-approximation queries alone.

\makeplot{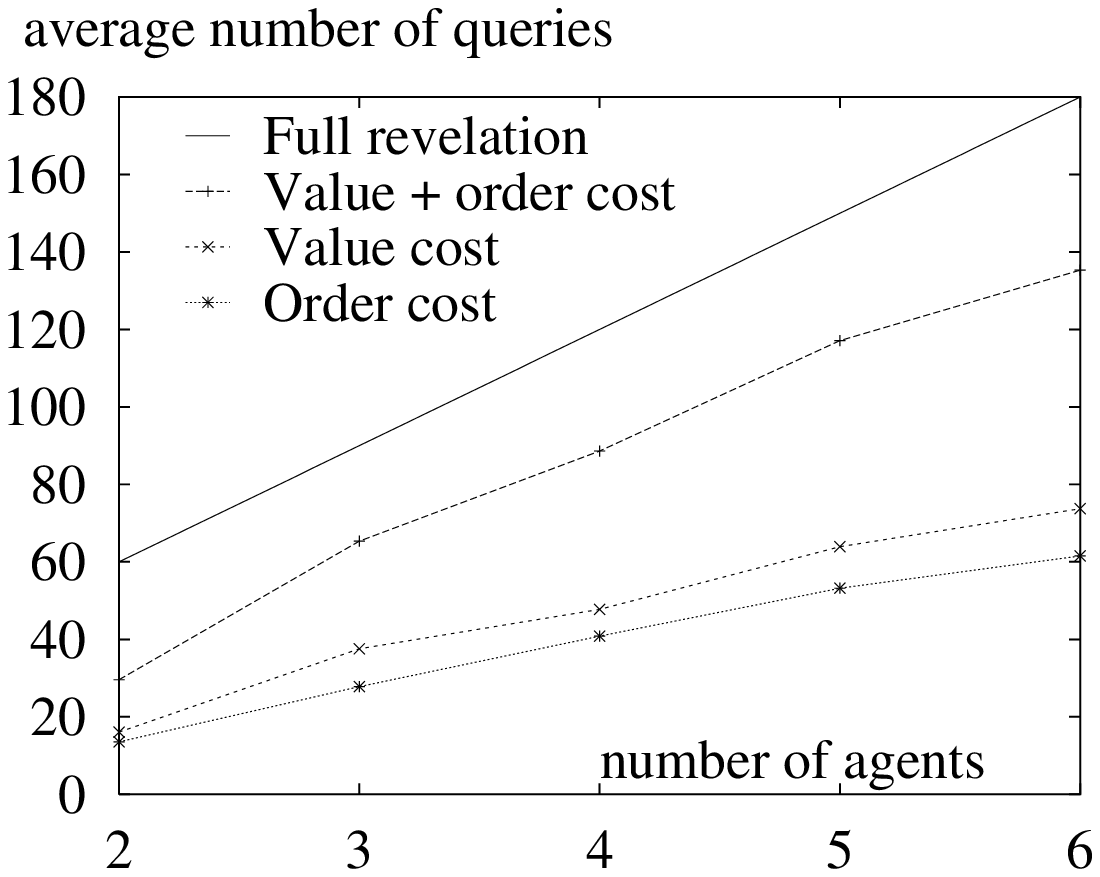}{Elicitation using bound-approximation and order queries.
The results are truncated to $k=9$ because a run at $k=10$ took longer
than 2 days.}

\mysection{Conclusions and future research}

In all of the elicitation schemes in this paper (except the
unrestricted random one), as the number of items for sale increases,
the amount of information elicited is a vanishing fraction of the
information collected in traditional ``direct revelation mechanisms''
where bidders reveal all their valuation information.  Each of the
elicitation schemes (except the rank lattice based one) also maintains
its benefit as the number of agents increases.

While the straightforward policies we analyzed work well,
some policies that attempt to be more intelligent actually perform
poorly (for example, the value query policy described by
Conen \& Sandholm~\cite{Conen01:Preference}, and some other policies we
tried).  These poorly-performing policies have in common that they use a
heuristic that maximizes the number of candidates that would be affected by
the query.  In contrast, the bound-approximation query policy that we
introduced benefits from the heuristic of maximizing the {\em change} in
bounds.  Future work includes designing additional useful heuristics for
selecting queries.

We showed theoretically that if it is possible to save revelation using an
elicitation policy, the simple unrestricted random elicitation policy saves
revelation.  We also presented theoretical and experimental results that
suggest that restricting value elicitation to allocatable bundles is
beneficial---which was assumed by Conen \&
Sandholm~\cite{Conen01:Preference} but is by no means obvious.  For the
other elicitation policies, our results were experimental.  Future work
includes studying their performance theoretically as well.

%
By using the {\em Clarke tax} mechanism~%
\cite{Clarke71:Multipart,Vickrey61,Groves73:Incentives,Varian95:Mechanism}
to determine the payments that the bidders have to pay, we can ensure that
in a Bayes-Nash equilibrium, each agent is motivated to answer the queries
truthfully, and is not less happy after the auction than before
it~\cite{Conen01:Preference} (under the usual assumption that the agents
have quasilinear preferences).  These payments can be computed by
determining an optimal allocation $n+1$ times: once overall, and once for
each agent removed in turn.  Even under the highly pessimistic assumption
that answers to queries in one of these problems do not help on the other
problems, determining the payments entails only an $n$-fold increase in the
number of queries.  Given that our results show that we have a
significantly better than $n$-fold benefit as the number of items grows,
this would not change the fact that only a vanishingly small fraction of
queries is asked.


Our bound-approximation queries take the incremental nature of
elicitation to a new level.  The agents are only asked for rough
bounds on valuations first, and more refined approximations are
elicited only on an as-needed basis.  A related approach would be to
propose a bound, and ask whether the agent's valuation is above or below
the bound.  This suggest a
relationship between preference elicitation and ascending combinatorial
auctions where the auction proceeds in rounds, and in each round the
bidders react to price feedback by revealing demand
(e.g.,~\cite{Parkes00:Iterative,Wurman00:AkBA,BDSV01,Parkes99:iBundle,Parkes00:Preventing}).


\subsubsection*{Acknowledgments}
The authors thank Maverick Woo for his help on the proofs of Propositions
\ref{prop:random_elicitor} and \ref{prop:bad_case_bound}.


\bibliographystyle{plain}
\bibliography{/afs/cs.cmu.edu/user/amem/refs/dairefs,./local}
\end{document}